\documentclass[12pt]{article}
\usepackage{amssymb,amsmath}
\usepackage{epsfig}
\newcommand{\sect}[1]{\setcounter{equation}{0}\section{#1}}

\def\N{{\mathcal N}}

\def\E{{\mathcal E}}
\def\S{{\mathcal S}}

\def\J{{\mathcal J}}

\def\ds{\displaystyle}
\def\r{\rho}
\def\a{\alpha}
\def\b{\beta}

\def\g{\gamma}

\def\s{\sigma}
\def\t{\tilde}

\def\n{\nu}
\def\k{\kappa}
\def\f{\phi}

\def\eps{\epsilon}
\def\l{\lambda}
\def\th{\theta}

\def\o{\omega}
\def\p{\partial}

\def\axs{AdS_5\times S^5}
\newcommand{\eq}[1]{\begin{equation} #1 \end{equation}}

\newcommand{\ml}[1]{\begin{multline} #1 \end{multline}}

\begin{document}

\begin{center}
{\bf{\Large Semiclassical strings in Lunin-Maldacena background }  \\
\vspace*{.35cm} }

\vspace*{1cm} N.P. Bobev${}^{\dagger}$, H. Dimov${}^{\ddag}$ and
R.C. Rashkov${}^{\dagger}$\footnote{e-mail:
rash@phys.uni-sofia.bg}

\ \\
${}^{\dagger}$Department of Physics, Sofia University, 1164 Sofia,
Bulgaria

\ \\

${}^{\ddag}$ Department of Mathematics, University of Chemical
Technology and Metallurgy, 1756 Sofia, Bulgaria

\end{center}

\vspace*{.8cm}

\begin{abstract}
The aim of this paper is to investigate semiclassical rotating
string configurations in the recently found Lunin-Maldacena
background. This background is conjectured to be dual to the
Leigh-Strassler $\b$-deformation of $\N=4$ SYM and therefore a
good laboratory for tests of the AdS/CFT correspondence beyond the
well explored $AdS_{5}\times S^{5}$ case. We consider different
multispin configurations of rotating strings by allowing the
strings to move in both the $AdS_{5}$ and the deformed $S^{5}$
part of the Lunin-Maldacena background. For all of these
configurations we compute the string energy in terms of the
angular momenta and the string winding numbers and thus provide
the possibility of reproducing our results from a computation of
the anomalous dimension of the corresponding dilatation operator.
This can be achieved by means of the Bethe ansatz techniques for
the relevant sectors of the corresponding Yang-Mills theory. We
also compare our results to those for multispin rotating strings
on $AdS_{5}\times S^{5}$.
\end{abstract}

\vspace*{.8cm}

\newpage

\sect{Introduction}
The idea for the correspondence between the
large N limit of gauge theories and string theory was proposed
over thirty years ago \cite{thooft} but its realization was given
when Maldacena conjectured the AdS/CFT correspondence
\cite{maldacena, witten,GKP}. Since then this became a major
research area and many fascinating discoveries were made in the
last years. These include the discovery of Gubser, Klebanov and
Polyakov \cite{gkp} that the energy of certain string
configurations in the limit of large quantum numbers reproduces
the behavior of the anomalous dimension of the corresponding SYM
operator. Soon after that Minahan and Zarembo \cite{mz} provided a
way to compute the anomalous dimension of a certain dilatation
operator by presenting it as the Hamiltonian of an integrable spin
chain. These two discoveries opened the door towards many
qualitative and quantative checks of the AdS/CFT beyond the
supergravity approximation. On the string side of the
correspondence many semiclassical string configurations on
$AdS_{5}\times S^{5}$ were studied and the classical energies as
well as their quantum corrections were obtained \footnote{see
\cite{ts3} and \cite{tseytlin} for a review}
\cite{ts}-\cite{ryang}. On the gauge theory side the anomalous
dimensions of the dilatation operator in certain sectors of the
$\N=4$ SYM were found (see \cite{beisphd} and references therein).
Moreover these dimensions coincide with the energies of the
corresponding semiclassical strings, thus providing remarkable
quantative proofs of the string theory/gauge theory correspondence
\cite{mz}-\cite{rashhri1}. Nevertheless the $AdS_{5}\times
S^{5}$/$\N=4$ SYM is the main and best explored example. Since the
$\N=4$ superconformal gauge theory is not appropriate
phenomologically we would like to extend the above ideas to less
supersymmetric Yang-Mills theories. Such attempts were made by
studying semiclassical strings in less supersymmetric backgrounds
(Maldacena-Nunez, Pilch-Warner and other confining and warped
geometries \cite{rash1}-\cite{pope}) but it was not completely
clear how to reproduce the string theory results from the SYM
side. Fortunately such a possibility emerged when Lunin and
Maldacena \cite{lunmal} found the gravity background dual to the
Leigh-Strassler $\b$-deformation of $\N=4$ SYM. A quantative check
of this correspondence for the $su(2)$ sector was made in
\cite{froits}, thus giving the hope that we can extend the
remarkable previous results to this case. Moreover in
\cite{frolov} the integrability of the string Hamiltonian on the
Lunin-Maldacena background was proven by finding a Lax pair. This
suggests that the interplay between integrable structures in
$AdS_{5}\times S^{5}$ \cite{bepolroi} and $\N=4$ SYM is also
present in this less supersymmetric case.

These recent developments give us the motivation to investigate
semiclassical strings on the Lunin-Maldacena background. In
\cite{froits} a simple two spin string ansatz was considered. In
this paper we would like to investigate multispin rotating string
solutions, having angular momenta both in the $AdS_{5}$ and the
$S^{5}$ part of the background. Since the anti-de Sitter piece of
the metric stays undeformed under the TsT transformation
generating the Lunin-Maldacena background we expect that the
string motion there will not differ from the $AdS_{5}\times S^{5}$
case. However the case of rotating string with three spins in the
deformed $S^{5}$ part will lead to some non-trivial results for
the energy. We will consider the following cases: i)three spins on
the deformed $S^{5}$ ($J_{1},J_{2},J_{3}$); ii)two spins on
$AdS_{5}$ and two spins on the deformed $S^{5}$
($S_{1},S_{2},J_{1},J_{2}$) and iii)the most general configuration
of five spins ($S_{1},S_{2},J_{1},J_{2},J_{3}$). In all of these
cases we compute the energy of the rotating string in terms of the
angular momenta and the string winding numbers. If we take the
limit $\t{\g}\rightarrow 0$ our results reproduce those found for
the undeformed $AdS_{5}\times S^{5}$ case \cite{arts}. This should
be expected since this is exactly the limit in which the
Lunin-Maldacena background reduces to the usual $AdS_{5}\times
S^{5}$. We would also like to note that it should be possible to
reproduce our results from the SYM side. For example the three
$S^{5}$ spin solution should be dual to operators from the $su(3)$
sector, their anomalous dimensions can be found by a spin chain
computation in the same spirit as this was done for the $su(2)$
case in \cite{froits}\footnote{After this paper was completed an interesting paper treating the three spin sector in the case of completely broken supersymmetry appeared \cite{frts}.}.

In the next section we will briefly review the form of the
Lunin-Maldacena background and find the energy for the rotating
three spin string ansatz. In section 3 the case of two spins on
$AdS_{5}$ and two spins on $S^{5}$ will be presented. In section 4
we will find the general rotating string solution with five spins
and compute its energy. Finally in the last section we will
present our conclusions and some open problems.

\sect{Three spin string solution in Lunin-Maldacena background}

We will begin by presenting the form of the supergravity
background found by Lunin and Maldacena: \eq{
ds^{2}_{str}=R^{2}\sqrt{H}\{ds^{2}_{AdS_{5}}+\sum_{i=1}^{3}(d\r_{i}^{2}+G\r_{i}^{2}d\f_{i}^{2})+(\t{\g}^{2}+\t{\s}^{2})G\r_{1}^{2}\r_{2}^{2}\r_{3}^{2}(\sum_{i=1}^{3}d\f_{i})^{2}\}
\label{1.1}} where \eq{\begin{array}{c}
\ds\frac{1}{G}=1+(\t{\g}^{2}+\t{\s}^{2})(\r_{1}^{2}\r_{2}^{2}+\r_{1}^{2}\r_{3}^{2}+\r_{2}^{2}\r_{3}^{2})
\qquad
H=1+\t{\s}^{2}(\r_{1}^{2}\r_{2}^{2}+\r_{1}^{2}\r_{3}^{2}+\r_{2}^{2}\r_{3}^{2})\\\\
B_{2}=R^{2}(\g Gw_{2}-\s w_{1}d\psi) \qquad\qquad
\psi=\ds\frac{\f_{1}+\f_{2}+\f_{3}}{3}\\\\
dw_{1}=\cos\a\sin^{3}\a\sin\th\cos\th d\a d\th \qquad
w_{2}=\r_{1}^{2}\r_{2}^{2}d\f_{1}d\f_{2}+\r_{2}^{2}\r_{3}^{2}d\f_{2}d\f_{3}+\r_{3}^{2}\r_{1}^{2}d\f_{3}d\f_{1}\\\\
\r_{1}=\sin\a\cos\th \qquad\qquad \r_{2}=\sin\a\sin\th
\qquad\qquad \r_{3}=\cos\a
\end{array} \label{1.2}}
Now we are going to write the Polyakov action for strings staying
at the center of $AdS_{5}$ and moving on the deformed five sphere
(i.e. strings on $R_{t}\times S^{5}_{\b}$).We will also suppose
that the deformation parameter is real ($\t{\g}\neq 0$,
$\t{\s}=0$).
 \ml{ S=-\ds\frac{R^{2}}{2}\int\frac{d\tau
d\s}{2\pi}[\g^{\a\b}\sqrt{H}(-\p_{\a}t\p_{\b}t+\p_{\a}\a\p_{\b}\a+\sin^{2}\a\p_{\a}\th\p_{\b}\th\\\\
+G\sin^{2}\a\cos^{2}\th\p_{\a}\f_{1}\p_{\b}\f_{1}+G\sin^{2}\a\sin^{2}\th\p_{\a}\f_{2}\p_{\b}\f_{2}+G\cos^{2}\a\p_{\a}\f_{3}\p_{\b}\f_{3}\\\\
+\t{\g}^{2}G\sin^{4}\a\cos^{2}\a\sin^{2}\th\cos^{2}\th(\p_{\a}\f_{1}+\p_{\a}\f_{2}+\p_{\a}\f_{3})(\p_{\b}\f_{1}+\p_{\b}\f_{2}+\p_{\b}\f_{3}))\\\\
-2\t{\g}
G\eps^{\a\b}(\sin^{4}\a\sin^{2}\th\cos^{2}\th\p_{\a}\f_{1}\p_{\b}\f_{2}+\sin^{2}\a\cos^{2}\a\sin^{2}\th\p_{\a}\f_{2}\p_{\b}\f_{3}\\\\+\sin^{2}\a\cos^{2}\a\cos^{2}\th\p_{\a}\f_{3}\p_{\b}\f_{1})]
 \label{1.3}}
Let us consider the following rotating string ansatz:
\eq{\f_{1}=\o_{1}\tau+m_{1}\s \qquad \f_{2}=\o_{2}\tau+m_{2}\s
\qquad \f_{3}=\o_{3}\tau+m_{3}\s  \label{1.4}} We impose also a
constant radii condition, namely $\a=\th=\ds\frac{\pi}{4}$ and the
global time is expressed through the world sheet time as
$t=\k\tau$. Then the equations of motion for $\a$ and $\th$ are
simply the following relations between the frequencies $\o_{i}$,
the winding numbers $m_{i}$ and the real deformation parameter
$\t{\g}$: \eq{
\o_{1}^{2}-m_{1}^{2}-\o_{2}^{2}+m_{2}^{2}+\ds\frac{\t{\g}}{2}(\o_{3}m_{2}-\o_{2}m_{3}-\o_{3}m_{1}+\o_{1}m_{3})=0
\label{1.5}} and \ml{
m_{1}^{2}-\o_{1}^{2}-\o_{2}^{2}+m_{2}^{2}+2\o_{3}^{2}-2m_{3}^{2}-\ds\frac{\t{\g}^{2}}{8}(\o_{1}+\o_{2}+\o_{3})^{2}+\ds\frac{\t{\g}^{2}}{8}(m_{1}+m_{2}+m_{3})^{2}\\\\+\t{\g}(\o_{2}m_{1}-\o_{1}m_{2})=0
\label{1.6}} We should also explore the Virasoro constraints which
in our case have the following form: \eq{
\o_{1}m_{1}+\o_{2}m_{2}+2\o_{3}m_{3}+\ds\frac{\t{\g}^{2}}{8}(\o_{1}m_{1}+\o_{2}m_{2}+\o_{3}m_{3}+\o_{1}m_{2}+\o_{1}m_{3}+\o_{2}m_{3})=0
\label{1.7}} and \ml{
\ds\frac{4\k^{2}}{G}=\o_{1}^{2}+\o_{2}^{2}+2\o_{3}^{2}+m_{1}^{2}+m_{2}^{2}+2m_{3}^{2}\\\\+\frac{\t{\g}^{2}}{8}(\o_{1}^{2}+\o_{2}^{2}+\o_{3}^{2}+2\o_{1}\o_{2}+2\o_{1}\o_{3}+2\o_{2}\o_{3}+m_{1}^{2}+m_{2}^{2}+m_{3}^{2}+2m_{1}m_{2}+2m_{1}m_{3}+2m_{2}m_{3})
\label{1.8}} It is easily seen that if \eq{m_{1}=m_{2}=-m_{3}=m
\qquad\qquad \o_{1}=\o_{2}=\o_{3}=\o \label{1.9}} (\ref{1.5}) and
(\ref{1.7}) are satisfied, from (\ref{1.6}) follows that $3\o=m$
and the second Virasoro constraint (\ref{1.8}) gives and
expression for $\k $ which is related to the energy in the
following way (we just note that $R^2=\sqrt{\l}$):
\eq{E=\sqrt{\l}\E=\ds\frac{R^{2}}{2\pi}\int_{0}^{2\pi}d\s\k=R^{2}\k
\label{1.10}} Before we calculate the energy let us first compute
the three conserved charges corresponding to the three angle
variables in our problem. \eq{\begin{array}{c}
\J_{1}=\ds\frac{J_1}{\sqrt{\l}}=\ds\frac{G}{2}(\o+\frac{3\t{\g}^{2}}{8}\o+\frac{3\t{\g}}{4}m)\\\\
\J_{2}=\ds\frac{J_2}{\sqrt{\l}}=\ds\frac{G}{2}(\o+\frac{3\t{\g}^{2}}{8}\o-\frac{3\t{\g}}{4}m)\\\\
\J_{3}=\ds\frac{J_3}{\sqrt{\l}}=G(\o+\ds\frac{3\t{\g}^{2}}{16}\o)
\end{array} \label{1.11}}
And thus for the full spin of our system we find:
\eq{\J=\J_{1}+\J_{2}+\J_{3}=2G\o+\ds\frac{9\t{\g}^{2}}{16}G\o=2G\o+\ds\frac{3\t{\g}^{2}}{16}Gm
\label{1.12}} The equation for $\k$ is simply: \eq{
\k=\sqrt{G(\o^{2}+m^{2}+\frac{9\t{\g}^{2}}{32}\o^{2}+\frac{\t{\g}^{2}}{32}m^{2})}
\label{1.13}} Thus for the energy we end up with the following,
messy at first sight,expression: \ml{
E=\sqrt{\l}\ds[\left(1+\frac{5\t{\g}^{2}}{16}\right)\frac{\J^{2}}{4}-\frac{3\t{\g}^{2}}{32}\left(1+\frac{9\t{\g}^{2}}{16}\right)m\J\\\\+\left(1+\frac{\t{\g}^{2}}{32}+\frac{9\t{\g}^{4}}{64(16+5\t{\g}^{2})}+\frac{81\t{\g}^{6}}{2056(16+5\t{\g}^{2})}\right)m^{2}]^{1/2}
\label{1.14}} Although it looks complicated this expression
reproduces the result for the energy of a three spin string in
pure $AdS_{5}\times S^{5}$ if we take the limit $\t{\g}\rightarrow
0$ (as should be expected), namely: \eq{
E=\sqrt{\l}\sqrt{\frac{\J^{2}}{4}+m^2} \label{1.15}}

\sect{Two spins in both $AdS_5$ and $S^5_{\t{\g}}$}

Here we will investigate semiclassical strings with two spins on
the $AdS_{5}$ part and two spins on the deformed $S^{5}$ part of
the Lunin-Maldacena background. The relevant Polyakov action is:
\ml{ S=-\ds\frac{R^{2}}{2}\int\frac{d\tau
d\s}{2\pi}[\g^{\a\b}\sqrt{H}(-\cosh^{2}\r\p_{\a}t\p_{\b}t+\p_{\a}\r\p_{\b}\r+\sinh^{2}\r\p_{\a}\psi\p_{\b}\psi\\\\+\sinh^{2}\r\cos^{2}\psi\p_{\a}\psi_{1}\p_{\b}\psi_{1}+\sinh^{2}\r\sin^{2}\psi\p_{\a}\psi_{2}\p_{\b}\psi_{2}+\p_{\a}\th\p_{\b}\th\\\\
+G\cos^{2}\th\p_{\a}\f_{1}\p_{\b}\f_{1}+G\sin^{2}\th\p_{\a}\f_{2}\p_{\b}\f_{2})-2\t{\g}
G\eps^{\a\b}\sin^{2}\th\cos^{2}\th\p_{\a}\f_{1}\p_{\b}\f_{2}]
 \label{2.1}}
We have imposed $\a=\pi/2$ and thus: \eq{
G^{-1}=1+\ds\frac{1}{4}(\t{\g}^{2}+\t{\s}^{2})\sin^{2}2\th
\qquad\qquad H=1+\ds\frac{\t{\s}^{2}}{4}\sin^{2}2\th \label{2.2}}
Now we will assume that $\t{\s}=0$, i.e. we will work in the real
deformed $AdS_{5}\times S^{5}$ background. It easily checked that
the following ansatz is compatible with the string equations of
motion: \eq{\begin{array}{c} t=\k\tau \qquad \r=const \qquad
\psi=\ds\frac{\pi}{4} \qquad \psi_{1}=\n_{1}\tau+n_{1}\s \qquad
\psi_{2}=\n_{2}\tau+n_{2}\s\\\\
\a=\ds\frac{\pi}{2} \qquad \th=\ds\frac{\pi}{4} \qquad
\f_{1}=\o_{1}\tau+m_{1}\s \qquad \f_{2}=\o_{2}\tau+m_{2}\s \qquad
\f_{3}=0
\end{array} \label{2.3}}
From the equations of motion for $\r$, $\psi$ and $\th$ follow
some relations between the winding numbers and the frequencies:
\eq{ \o_{1}^{2}-m_{1}^{2}=\o_{2}^{2}-m_{2}^{2} \qquad
\n_{1}^{2}-n_{1}^{2}=\n_{2}^{2}-n_{2}^{2} \qquad
\k^{2}=\n_{1}^{2}-n_{1}^{2} \label{2.4}} The Virasoro constraints
of our system adopt the following form: \eq{\begin{array}{c}
\ds\frac{\sinh^{2}\r}{2}(\n_{1}n_{1}+\n_{2}n_{2})+\ds\frac{G}{2}(\o_{1}m_{1}+\o_{2}m_{2})=0\\\\
\k^{2}\cosh^{2}\r=\ds\frac{\sinh^{2}\r}{2}(\n_{1}^{2}+\n_{2}^{2}+n_{1}^{2}+n_{2}^{2})+\ds\frac{G}{2}(\o_{1}^{2}+\o_{2}^{2}+m_{1}^{2}+m_{2}^{2})
\end{array}\label{2.5}}
The equations of motion and the first Virasoro constraint are
satisfied if we choose: \eq{ \n=\n_{1}=\n_{2} \qquad
\o=\o_{1}=\o_{2} \qquad n=n_{1}=-n_{2} \qquad m=m_{1}=-m_{2}
\label{2.6}} For the two $AdS_{5}$ angular momenta ($\S_{1}$,
$\S_{2}$) and the two $S^{5}$ angular momenta ($\J_{1}$, $\J_{2}$)
we obtain: \eq{ \S_{1}=\S_{2}=\ds\frac{\sinh^{2}\r}{2}\n \qquad
\J_{1}=\J_{2}=\ds\frac{G}{2}\o-\ds\frac{\t{\g} G}{4}m \label{2.7}}
The full $AdS_{5}$ and $S^{5}$ angular momenta are simply: \eq{
\S=\S_{1}+\S_{2}=\sinh^{2}\r\n \qquad
\J=\J_{1}+\J_{2}=G\o-\ds\frac{\t{\g} G}{2}m \label{2.8}} This
leads to the following relation: \eq{
\o=\J+\ds\frac{\t{\g}}{2}(m+\ds\frac{\t{\g}}{2}\J) \label{2.9}}
And thus the second Virasoro constraint can be expressed as: \eq{
\k^{2}\cosh^{2}\r=\sinh^{2}\r(\n^{2}+n^{2})+\J^{2}+(m+\ds\frac{\t{\g}}{2}\J)^{2}
\label{2.10}} The full energy of our system is $\E=\k\cosh^{2}\r$
and we find the following relation: \eq{
\ds\frac{\E}{\k}-\ds\frac{\S}{\n}=1 \label{2.11}} Using this
relation, the second Virasoro constraint and the relation coming
from the equation of motion for $\r$ we end up with: \eq{
2\E\k-\k^{2}=2\sqrt{n^{2}+\k^{2}}\S+\J^{2}+(m+\ds\frac{\t{\g}}{2}\J)^{2}
\label{2.12}} This expression reduces to the analogous expression
found by Arutyunov, Russo and Tseytlin \cite{arts} if we take the
limit $\t{\g}\rightarrow 0$. This is exactly what one should
expect because this limit reproduces the well known $AdS_{5}\times
S^{5}$ background, which is the case considered in \cite{arts}.

\sect{Generalized rotating string}

Considering the results from the preceding sections it seems
natural to combine them in order to investigate the most general
rotating string ansatz. This is the case of two spins in the
$AdS_{5}$ part and three spins in the $S^{5}$ part of the geometry
($S_{1},S_{2},J_{1},J_{2},J_{3}$). Following the same procedure we
start with the relevant string action:

\ml{ S=-\ds\frac{R^{2}}{2}\int\frac{d\tau
d\s}{2\pi}[\g^{\a\b}\sqrt{H}(-\cosh^{2}\r\p_{\a}t\p_{\b}t+\p_{\a}\r\p_{\b}\r+\sinh^{2}\r\p_{\a}\psi\p_{\b}\psi\\\\+\sinh^{2}\r\cos^{2}\psi\p_{\a}\psi_{1}\p_{\b}\psi_{1}+\sinh^{2}\r\sin^{2}\psi\p_{\a}\psi_{2}\p_{\b}\psi_{2}+\p_{\a}\a\p_{\b}\a+\sin^{2}\a\p_{\a}\th\p_{\b}\th\\\\
+G\sin^{2}\a\cos^{2}\th\p_{\a}\f_{1}\p_{\b}\f_{1}+G\sin^{2}\a\sin^{2}\th\p_{\a}\f_{2}\p_{\b}\f_{2}+G\cos^{2}\a\p_{\a}\f_{3}\p_{\b}\f_{3}\\\\
+\t{\g}^{2}G\sin^{4}\a\cos^{2}\a\sin^{2}\th\cos^{2}\th(\p_{\a}\f_{1}+\p_{\a}\f_{2}+\p_{\a}\f_{3})(\p_{\b}\f_{1}+\p_{\b}\f_{2}+\p_{\b}\f_{3}))\\\\
-2\t{\g}
G\eps^{\a\b}(\sin^{4}\a\sin^{2}\th\cos^{2}\th\p_{\a}\f_{1}\p_{\b}\f_{2}+\sin^{2}\a\cos^{2}\a\sin^{2}\th\p_{\a}\f_{2}\p_{\b}\f_{3}\\\\+\sin^{2}\a\cos^{2}\a\cos^{2}\th\p_{\a}\f_{3}\p_{\b}\f_{1})]
 \label{1.3}}
We will again examine the case of real deformation parameter and
use the following ansatz:
\eq{\begin{array}{c} t=\k\tau \qquad
\r=const \qquad \psi=\ds\frac{\pi}{4} \qquad
\psi_{1}=\n_{1}\tau+n_{1}\s \qquad
\psi_{2}=\n_{2}\tau+n_{2}\s\\\\
\a=\ds\frac{\pi}{4} \qquad \th=\ds\frac{\pi}{4} \qquad
\f_{1}=\o_{1}\tau+m_{1}\s \qquad \f_{2}=\o_{2}\tau+m_{2}\s \qquad
\f_{3}=\o_3\tau+m_3\s
\end{array} \label{3.1}}
From the equations of motion for $\r$, $\psi$, $\a$ and $\th$ we
can extract the following relations between the frequencies and
the winding numbers.
\eq{\begin{array}{c}
\n_{1}^{2}-n_{1}^{2}=\n_{2}^{2}-n_{2}^{2}
\qquad \k^{2}=\n_{1}^{2}-n_{1}^{2}\\\\
\o_{1}^{2}-m_{1}^{2}-\o_{2}^{2}+m_{2}^{2}+\ds\frac{\t{\g}}{2}(\o_{3}m_{2}-\o_{2}m_{3}-\o_{3}m_{1}+\o_{1}m_{3})=0\\\\
m_{1}^{2}-\o_{1}^{2}-\o_{2}^{2}+m_{2}^{2}+2\o_{3}^{2}-2m_{3}^{2}-\ds\frac{\t{\g}^{2}}{8}(\o_{1}+\o_{2}+\o_{3})^{2}\\\\+\ds\frac{\t{\g}^{2}}{8}(m_{1}+m_{2}+m_{3})^{2}+
\t{\g}(\o_{2}m_{1}-\o_{1}m_{2})=0
\end{array} \label{3.2} }
We should also impose the Virasoro constraints, they will provide
one more relation between the parameters and an equation for $\k$.
We will retain from presenting their explicit form here but he
expression are analogous to the previous two cases. In order to
satisfy the constraints and the equations of motion we should
choose: \eq{ \n=\n_{1}=\n_{2} \qquad \o=\o_{1}=\o_{2}=\o_3 \qquad
n=n_{1}=-n_{2} \qquad m=m_{1}=m_{2}=-m_3 \label{3.3}} It is again
straightforward to compute the angular momenta in both parts of
the background:
\eq{\begin{array}{c}
\J=\J_{1}+\J_{2}+\J_{3}=2G\o+\ds\frac{9\t{\g}^{2}}{16}G\o=2G\o+\ds\frac{3\t{\g}^{2}}{16}Gm\\\\
\S=\S_{1}+\S_{2}=\sinh^{2}\r\n \end{array}\label{3.4}} We just
remind that in this case $G^{-1}=1+\frac{5}{16}\t{\g}^2$. The
equation for $\k$ following from the second Virasoro constraint
is: \eq{
\k^{2}\cosh^{2}\r=\sinh^{2}\r(\n^{2}+n^{2})+G(\o^{2}+m^{2}+\frac{9\t{\g}^{2}}{32}\o^{2}+\frac{\t{\g}^{2}}{32}m^{2})
\label{3.5}} The energy of the rotating string is $\E=\k\cosh^2\r$
and it is related to the $AdS$ angular momentum as: \eq{
\ds\frac{\E}{\k}-\ds\frac{\S}{\n}=1 \label{3.6}} And we can
extract an analogous to (\ref{2.12}) expression:
 \ml{
2\E\k-\k^{2}=2\sqrt{n^{2}+\k^{2}}\S+\ds\left(1+\frac{5\t{\g}^{2}}{16}\right)\frac{\J^{2}}{4}-\frac{3\t{\g}^{2}}{32}\left(1+\frac{9\t{\g}^{2}}{16}\right)m\J\\\\+\left(1+\frac{\t{\g}^{2}}{32}+\frac{9\t{\g}^{4}}{64(16+5\t{\g}^{2})}+\frac{81\t{\g}^{6}}{2056(16+5\t{\g}^{2})}\right)m^{2}
\label{3.7}} This expression again reduces to the one known from
the undeformed $\axs$ case if we take the limit $\t{\g}\rightarrow
0$.

\sect{Conclusions and Outlook}

In this paper we have studied rotating strings configurations in
the recently found Lunin-Maldacena background. These semiclassical
strings are an important tool for proving the AdS/CFT beyond the
supergravity approximation. We have found the energy of different
rotating strings in terms of the angular momenta and the string
winding numbers. In the limit of zero deformation parameter we
reproduce the well known results from the $\axs$ case, this should
be expected because this is the limit in which the Lunin-Maldacena
background reduces to the usual $\axs$.

Our work can be extended in several ways \footnote{For other related work on the subject see \cite{koch}-\cite{nunez} }. First of all we can work
with the most general ansatz with different frequencies and
winding numbers and find the energy behavior in terms of the
angular momenta. It is also important to consider fluctuations
around this classical solutions. They will provide the corrections
to the above found energies and also clarify the stability of
these solutions. Maybe the most important open problem is to
reproduce our results from the gauge theory side. We believe that
this could be done on the level of effective actions \cite{krucz}.
This was the approach in \cite{froits}, where an exact agreement
was found for the $su(2)$ sector and we expect that this agreement
could be extended for the $su(3)$ sector considered in the second
section.

There is one more interesting class of semiclassical strings -
pulsating strings. It is worth investigating such pulsating
solutions in the Lunin-Maldacena background and see how the
deformation affects the form of the solution. We plan to address
this issue in a future paper.

\textbf{Acknowledgments} We would like to thank Carolos Nunez for useful suggestions and comments. The work of N.P.B. was supported by an
EVRIKA foundation educational award.

\end{document}